\documentclass[11pt,twocolumn,english,aps,preprint,showpacs]{revtex4}
\usepackage{comment}
\usepackage{epsfig}
\usepackage{amsmath}
\usepackage{dcolumn}

\newcommand{\te}[1]{\text{#1}}

\begin{document}

\title{ Higher order effects in the $^{16}\te{O}(d,p)^{17}\te{O}$ and
$^{16}\te{O}(d,n)^{17}\te{F}$ transfer reactions
\thanks{Aux\'{\i}lio FAPESP no.2001/06676-9}}

\author{M. Assun\c{c}\~ao}
\author{R. Lichtenth\"aler}
\author{V. Guimar\~aes}
\author{A. L\'epine-Szily}
\author{G. F. Lima} 
\affiliation{Departamento de F\'{\i}sica Nuclear,
 Instituto de F\'{\i}sica da Universidade de S\~ao Paulo,
        CP 66318, 05315-970 S\~ao Paulo SP, Brasil}
\author{A. M. Moro}
\affiliation{Departamento de
F\'{\i}sica, Instituto Superior T\'ecnico, Taguspark,
2780-990 Porto Salvo, Portugal   }
\affiliation{Departamento
de FAMN,
Universidad de Sevilla,
Apdo. 1065, E-41080 Sevilla, Spain}

\begin{abstract}

Full Coupled Channels Calculations were performed for the $^{16}\text{O}(d,n)^{17}\text{F}$ and 
$^{16}\text{O}(d,p)^{17}\text{O}$ transfer reactions at several deuteron 
incident energies from $E_{lab}=2.29$ MeV up to $3.27$ MeV.
A strong polarization effect between the entrance channel and  the 
transfer channels $^{16}\text{O}(d,n)^{17}\text{F}(1/2^{+},0.495)$ and 
$^{16}\text{O}(d,p)^{17}\text{O}(1/2^{+},0.87)$ 
was observed. 
This polarization effect had to be taken into account 
in order to obtain realistic spectroscopic factors from these reactions.   
\vspace{1pc}
\end{abstract}
\date{\today}
\maketitle

\section{Introduction}

The interest in the experimental and theoretical study of few nucleon 
transfer reactions has been renewed in the last 
years mainly due to the possibility to obtain information of astrophysical 
relevance from these reactions
\cite{Xu94,Muk97,Gag99,Fer00}. Direct measurement of  
capture reactions  at energies of 
astrophysical interest is, in some cases, nearly impossible 
due to the low 
reaction yield, especially if the capture involves exotic nuclei. 
Alternative indirect methods, such as the asymptotic normalization
coefficient (ANC) method, based on the analysis of 
breakup \cite{Tra01} or transfer reactions \cite{Xu94}, have been 
used as a tool to obtain astrophysical $S$-factors. 
The advantage of  indirect approaches comes from the fact 
that transfer and breakup reactions can be measured at higher 
energies, where the cross sections are much larger. However, to obtain useful 
information from transfer reactions one needs to understand, as clear 
as possible, the reaction mechanism involved. 

Actually, by comparing the  DWBA calculations with the experimental angular 
distributions it is possible to determine  the
spectroscopic factors of the transferred particles
in the target and projectile system. However, as a first 
order theory, the DWBA method is based on the assumption that 
the transfer occurs in 
one single step from the ground state of the entrance channel directly 
to one specified state of the final nucleus in the outgoing channel. 
Within the DWBA, the transfer is proportional to the 
product of the  spectroscopic factors of the transfered 
particle in the projectile and target. So, if one of the 
spectroscopic amplitudes 
is known, the other can be obtained by comparing the DWBA calculation 
with the experimental angular distribution. 
Spectroscopic factors extracted from transfer 
reaction analyses, appear to be in some cases energy dependent, 
indicating that a simple 
DWBA analysis may not be applicable. Also, if one of the nuclei in 
the entrance channel is strongly excited
during the collision, the one channel approach implicit in the 
DWBA scheme might
be inappropriate. 
In this case, the Coupled Channels Born 
Approximation (CCBA) approach is more suitable  
\cite{Nun01}. 
In the CCBA formalism, the transfer is still considered as a 
one step process but the effect of the coupling to a set of selected 
excited states 
of the projectile or target are included 
explicitly.  The spectroscopic amplitudes 
obtained in CCBA  will be the result of the mixing of amplitudes 
for different excited states. Due to this mixing,
the results of such calculation can not be used to extract the asymptotic
normalization coefficient  for astrophysical calculations. 
In addition to the coupling to inelastic excitations,   
other effects, such as strong polarization
between the entrance channel  and the transfer 
channels, might be important to describe the data. 
In this case, multistep transfer going forward and backward  
between   states of different partitions could give rise to a rearrangement 
of the flux of the specified channels and the Coupled Reaction Channels (CRC) 
formalism should be used instead. 
Although in the CRC formalism the final cross section of the  transfer 
channel will be affected by this polarization, it 
may be still possible to obtain the ANC coefficients and
$S$-factors provided that the coupling with other intermediate excited 
states are negligible.  In case of  weak coupling between excited 
states and strong polarization, only one spectroscopic amplitude 
is involved and it can be reliably extracted for 
astrophysics purposes.

In this paper, we investigate the importance of considering channel 
couplings effects in the analysis of the 
$^{16}\text{O}(d,n)^{17}\text{F}$ and $^{16}\text{O}(d,p)^{17}\text{O}$ 
transfer reactions, at incident deuteron energies from 
$E_d$=2.279~MeV to $E_d$=3.155 MeV, for 
which experimental data exist \cite{Die68}.
By performing CRC calculations, we show that 
if realistic spectroscopic information is to be obtained from these 
reactions one has to go beyond the Born approximation.

The paper is organized as follows. In section \ref{sec:DWBA}, DWBA and
CCBA calculations are presented for the reactions under study. In section
\ref{sec:CRC}, CRC calculations are performed for the same reactions. The 
results
obtained with the different reaction formalisms are discussed in section 
\ref{sec:discussion}.  Finally, in section \ref{sec:summary} we summarize 
the main conclusions  achieved in this work.

\begin{figure}
{\centering \resizebox*{0.95\columnwidth}{!}
{\includegraphics{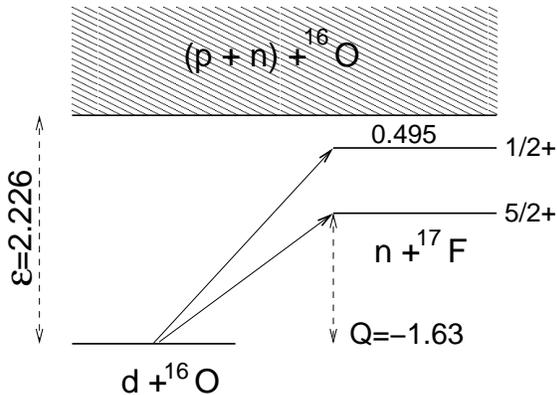}} \par}
\caption{\label{Fig:o16dn_coup} Coupling scheme for the 
 $^{16}\text{O}(d,n)^{17}\text{F}$ reaction. Solid arrows indicate 
transitions considered
in the DWBA calculations.}
\end{figure}

\section{DWBA analysis}\label{sec:DWBA}

The $^{16}\text{O}(d,n)^{17}\text{F}$ transfer reaction was analyzed
in terms of the DWBA formalism which, in prior form, involves the transition
operator $V_\text{[p-$^{16}$O]}+U_{[n-^{16}\text{O}]}-U_{[d-^{16}\text{O}]}$. 
The distorting potential for the entrance channel, $U_{[d-^{16}\text{O}]}$,
was considered as a variation of the Satchler parameterization \cite{Sat66}. A
slight modification of the parameters was introduced in order to improve the 
fit to the data. The exit channel optical potential, $U_{[n-^{17}\text{F}]}$,
was determined from 
the survey of Rosen \cite{Rosen66}. This global parameterization was also 
used for the core-core
interaction, $U_{[n-^{16}\text{O}]}$, although only the real part of the
potential was retained.
These potentials are listed in Table \ref{Tab:OP1}.
For the binding potential of the $^{17}$F nucleus a
Woods-Saxon form with the standard parameters $r_{0}$=1.25~fm
and $a_{0}$=0.65~fm was considered. The valence proton in the 
ground state of $^{17}$F is assumed to occupy the $1d_{5/2}$ orbit, with a 
spectroscopic factor adjusted to reproduce the experimental 
angular distribution data. 
For the $p$-$n$ binding potential, $V_{pn}$, a Gaussian form 
$V_{pn}(r)=-v_{0}\exp (r^{2}/a^{2})$
with $a=1.484$~fm and $v_{0}$=72.15~MeV was used. These parameters
were chosen to reproduce the r.m.s. and binding energy of the deuteron. 
A pictorial representation of this reaction is shown in 
Fig.~\ref{Fig:o16dn_coup}. The DWBA transitions considered in our calculations
are indicated  by solid arrows.

\begin{table*}
\caption{\label{Tab:OP1}Optical model parameters used in the 
DWBA calculations. All
potentials have a Woods-Saxon derivative imaginary potential.}
\begin{ruledtabular}
\begin{tabular}{ccccccccccc}
 
System&
 $V_{0}$&
 $r_{0}$&
 $a_{0}$&
 $W$$_{d}$&
 $r_{i}$&
 $a_{i}$&
 $V_{so}$&
 $r_{so}$&
 $a_{so}$&
Ref.\\
\hline 
$d+^{16}$O (a) & 110.0  & 1.012  & 0.876  & 9.3    & 1.837  & 0.356 & 6.0 & 1.4 & 0.7 &
\cite{Sat66}\\
$n+^{17}$F (b) &
-49.3+0.33$E_{c.m.}$ & 1.25 & 0.65 & 5.75  & 1.25 & 0.70  &  5.5 & 1.25 & 0.65 &
\cite{Rosen66}\\
$p+^{17}$O (c) &
-53.8+0.33$E_{c.m.}$&
1.25&
0.65&
7.5&
1.25&
0.70&
5.5&
1.25&
0.65&
\cite{Rosen66}\\
$n+^{17}$F (d) &
-&
-&
-&
65&
2.0&
0.332&
-&
-&
-&
(DPP)\\
$p+^{17}$O (e) &
-11.1&
1.25&
0.58&
2.3&
1.25&
1.07&
-&
-&
-&
(DPP)\\

\end{tabular}
\end{ruledtabular}
\end{table*}

\begin{figure}
{\centering \resizebox*{0.95\columnwidth}{!}
{\includegraphics{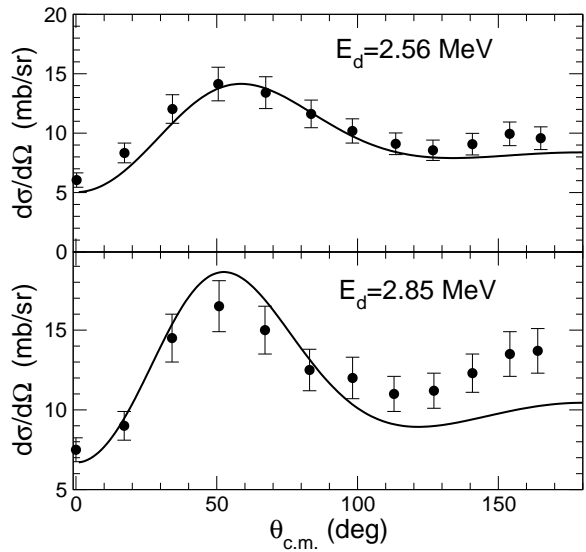}} \par}
\caption{\label{Fig:o16dn0_dwba}DWBA calculations for the proton transfer
reaction $^{16}\text{O}(d,n_{0})^{17}\text{F}$ at $E_{d}$=2.56~MeV
and $E_{d}$=2.85~MeV. In both cases, a spectroscopic
factor of 0.85 is used for the $\langle$$^{17}\text{F}|^{16}\text{O}\rangle$ 
overlap.}
\end{figure}

\begin{figure}
{\centering \resizebox*{0.95\columnwidth}{!}{\includegraphics{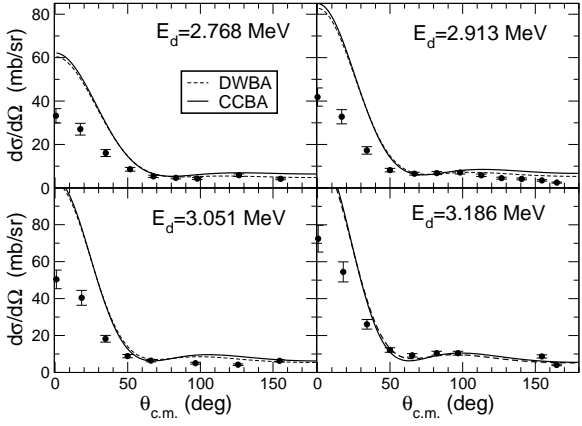}} \par}
\caption{\label{Fig:o16dn1_dwba}DWBA and CCBA (prior form) calculations 
for the proton
transfer reaction $^{16}\text{O}(d,n_{1})^{17}\text{F}$(0.495)
at four different scattering energies. A spectroscopic factor of 1
was used for the $\langle^{17}\text{F}|^{16}\text{O}\rangle$ overlap.}
\end{figure}

In Fig.~\ref{Fig:o16dn0_dwba}, we present the 
DWBA-prior calculations
for $^{16}\text{O}(d,n_{0})^{17}\text{F}_{gs}$ reaction, 
at two different scattering energies, along with 
the experimental angular distribution from Ref.~\cite{Die68}.
To separate the direct cross section from the compound nucleus
component we have considered energy averaged angular distributions.
The average compound nucleus (CN)
contributions for this reaction, estimated by 
Dietzsch {\it et al.} \cite{Die68} are 1.5 mb/sr
for $E_d$=2.56 MeV and 2.0 mb/sr for $E_d$=2.85 MeV. Since 
this contribution is roughly
angular independent, we just added these values to the calculated 
angular distributions.   The spectroscopic 
factors $S$=0.85, for the $\langle^{17}\text{F}|^{16}\text{O}\rangle$ 
vertex, and 
$S$=1, for $\langle d|n\rangle$,
were used at both incident energies. The overall agreement between the 
calculated and experimental angular distributions, in both
shape and normalization, is good, although the calculations underestimate
the experimental data at the larger angles for $E_{d}$=2.85~MeV. Similar
DWBA calculations
were performed for the proton transfer to the first excited state in $^{17}$F
at $E_{x}$=0.495~MeV. This state has $J^{\pi}$=1/2$^+$ 
assignment, which corresponds mainly to a $2s_{1/2}$ valence proton
coupled to a zero-spin $^{16}$O core. In Fig.~\ref{Fig:o16dn1_dwba},
the DWBA calculations (dashed lines) for the 
$^{16}\text{O}(d,n_{1})^{17}\text{F}$(0.495) reaction, are compared 
with the experimental
angular distributions obtained at four different incident energies. 
As can be seen in the figure, the calculations
overestimate the data for all energies considered. A spectroscopic
factor of the order of $S\approx 0.7$ for 
the $\langle^{17}\text{F}^{*}|^{16}\text{O}\rangle$ 
overlap would be required to reproduce
the data. This small value is in clear disagreement with 
shell model calculations 
and with previous measurements at higher energies (8-12 MeV) 
\cite{Tho69,Oli69,Lew99},
which give spectroscopic factor close to 1 for 
this overlap.

\begin{figure}
{\centering \resizebox*{0.97\columnwidth}{!}{\includegraphics{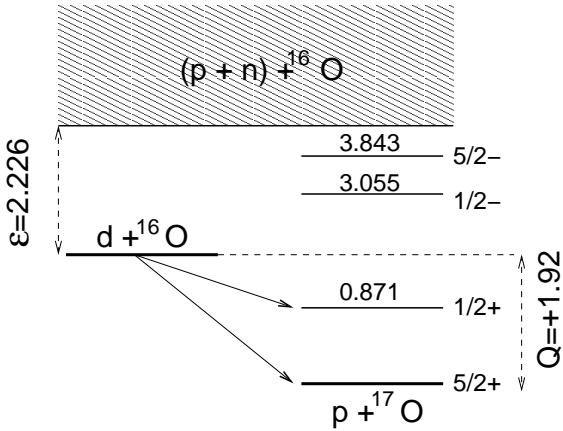}} \par}
\caption{\label{Fig:o16dp_coup} Coupling scheme for the  
$^{16}\text{O}(d,p)^{17}\text{O}$
reaction}
\end{figure}

\begin{figure}
{\centering \resizebox*{0.95\columnwidth}{!}{\includegraphics{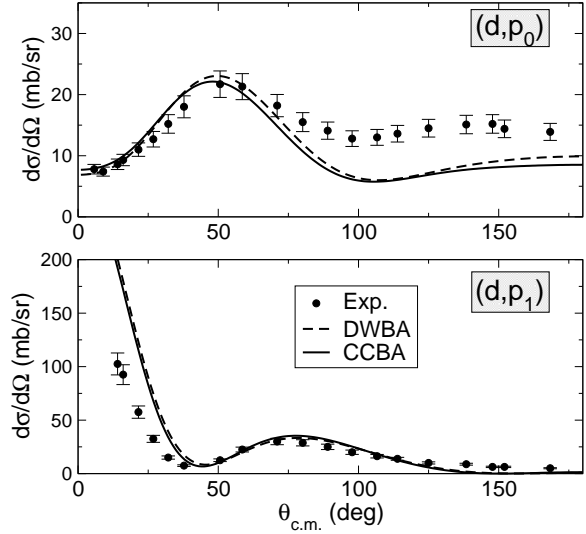}} \par}
\caption{\label{Fig:o16dp_dwba}DWBA and CCBA (prior) 
calculations for the neutron
transfer reaction $^{16}\text{O}(d,p)^{17}\text{F}$
at $E_{d}$=2.85~MeV. }
\end{figure}

Similar calculations were performed for the neutron transfer reaction 
$^{16}\text{O}(d,p)^{17}\text{O}$. The potentials
used in this case are the same as those used in the 
analysis of the proton transfer reaction, except
for Rosen potential, which predict slightly different potentials
for protons and neutrons (see potential (e) in Table \ref{Tab:OP1}).  The 
ground ($5/2^{+}$) 
and first excited ($1/2^{+}$, $E_x$=0.871 MeV) states in $^{17}$O 
were considered in the analysis (see scheme in Fig.~\ref{Fig:o16dp_coup}).
Pure single-particle configurations for the valence neutron, with spectroscopic
factors 0.85 and 1, for the ground and excited states, respectively,
were assumed. The calculated angular distributions for the
deuteron incident energy
$E_{d}$=2.85~MeV are presented in Fig.~\ref{Fig:o16dp_dwba}.
As in the case of the $(d,n)$ reaction, there is a good agreement between the
calculated and experimental angular distribution for the
$(d,p_{0})$ reaction at forward angles, while a clear overestimation for the
$(d,p_{1})$ data is observed. A spectroscopic factor of about
0.6 would be required to fit the forward angle data which, as in
the $^{17}$F case, is not consistent with the marked single-particle
character expected for this state.

\subsection*{CCBA calculations}

\begin{figure}
{\centering \resizebox*{0.95\columnwidth}{!}{\includegraphics{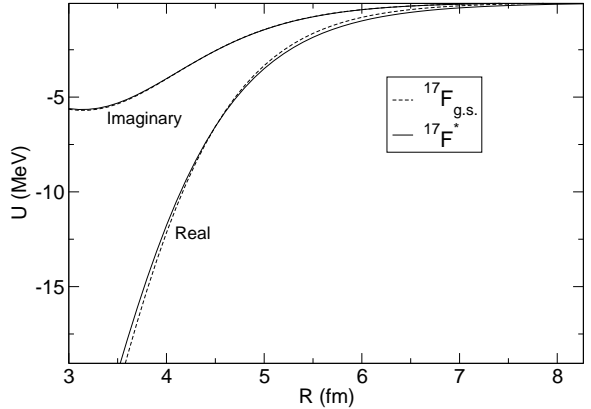}} \par}
\caption{\label{Fig:Ufold}Diagonal part of the cluster-folded potential for
$n+^{17}\text{F}_{gs}$ and $n+^{17}\text{F}^{*}$.}
\end{figure}

One of the main sources of ambiguity in the DWBA calculations presented
above is the optical potential for the exit channel ($n+^{17}$F
or $p+^{17}$O). In the former case, the weakly bound
nature of the $^{17}$F($1/2^+$,0.495) state produces a long tail in the
bound state wave function, making the system more  diffuse than the
ground state. For instance, the r.m.s., calculated in a Woods-Saxon
well with standard parameters, are 3.75~fm
and 5.40~fm for the ground and excited state, respectively. 
Therefore, the different character of these two states might cast 
doubt on the validity of the global (Rosen \cite{Rosen66})
parameterization
used to describe the ${n+}^{17}$F$^*$ elastic scattering. 
In addition, couplings between the ground 
and excited states of $^{17}$F are neglected in the DWBA calculations.
These effects can be properly taken into account within the CCBA formalism 
\cite{Sat83}, where the final state wavefunction is obtained
as a solution of the set of coupled equations, where diagonal as well
as non-diagonal couplings between a set of selected projectile
or target states are considered.
In the case under consideration, 
these couplings can be naturally generated by assuming that the 
$^{16}$O behaves as an inert core and folding the 
$p$-$n$ and $n$-$^{16}$O interactions, i.e.,
\begin{equation}
\label{eq:ccpot}
U_{ij}=\langle \phi _{i}|V_\text{p-n}+U_\text{n-$^{16}$O}|\phi _{j}\rangle 
\end{equation}
where $i$ and $j$ refer to either the ground or the excited state.
In our calculation, only the ground and first excited states in $^{17}$F
were considered in the model space (see Fig.~\ref{Fig:o16dn_coup}).   
Note that the effect of the weak binding energy of this excited state 
is implicitly included in
the inter-cluster wave-function $\phi _{i}$. The resulting diagonal
potentials for both states are shown in Fig.~\ref{Fig:Ufold}. 
The weakly bound nature of the excited state  produces a  slightly 
more diffuse real potential.
Note that the imaginary potentials are almost identical. 
The small difference between these two folded potentials 
indicates that the halo
effect does not show up in the folded potential.
This conclusion is confirmed  in Fig.~\ref{Fig:o16dn1_dwba}, where  
CCBA
calculations (solid lines) are compared with  DWBA calculations 
(dashed lines) for the $(d,n)$ channel. As one can
see, the calculated angular distributions are very similar in both
approaches, indicating that final state interactions 
arising from target excitation plays a negligible role in this reaction.

\section{CRC calculations}\label{sec:CRC}

In the previous sections we have shown that both DWBA and CCBA calculations
do not  reproduce adequately the transfer cross section for the 
$^{16}\text{O}(d,n_{1})^{17}\text{F}$ and $^{16}\text{O}(d,p_1)^{17}\text{O}$ 
reactions, unless very
small spectroscopic factors are used for the 
$\langle^{17}\text{F}^{*}|^{16}\text{O}\rangle$
and $\langle^{17}\text{O}^{*}|^{16}\text{O}\rangle$ overlap wave functions. 
The accuracy of these two approaches rely on the  validity
of the Born approximation (BA). 
In this section we assess the accuracy of the BA for
the present reaction by performing Coupled Reaction Channels (CRC)
calculations. 

In the CRC approach \cite{Sat83}, the optical potentials
for the entrance and exit channels must be understood as 
{\it bare} potentials. Once the transfer couplings are 
set in both directions, the bare potentials are defined as to 
reproduce the elastic channel on their respective channels.
In principle, all the parameters of the potentials involved in this 
treatment could be considered as free parameters. These parameters could 
be simultaneously determined in the optimization of the overall 
agreement between the calculated cross section and the  
experimental angular distributions for 
the $(d,d)$, $(d,p)$ and $(d,n)$ channels.
In addition, the spectroscopic factors can also be treated
as adjustable parameters. However, fitting all optical potentials and
spectroscopic factors simultaneously  would turn the  
searching procedure very lengthy. 
Moreover, the lack of experimental data for the
proton and neutron elastic scattering scattering for the exit channels, 
makes it hard to determine realistic OP for these systems. Consequently, 
the OP parameters for these exit channels 
were kept fixed to the values given by the Rosen parameterization.
We verified nevertheless that slight changes in these parameters, 
within physically reasonable
constraints, did not affect significantly 
the agreement between the calculation and the data nor 
the extracted spectroscopic factors. 
We found also that, in order to obtain a good description of the elastic 
cross section of the entrance channel, the spin-orbit term in the 
deuteron optical potential had to be  eliminated.
The initial parameters of these potentials were the same 
as those used in the  
DWBA calculations described in the previous sections. 
The non-orthogonality correction \cite{Ima74} was also included  
in the CRC calculations,
since this effect was found to be important in all cases considered. 
All calculations were
performed with the search routine of the 
computer code FRESCO \cite{Thom88}, version frxy.

The estimates of the CN contributions obtained in Ref.~\cite{Die68} rely 
to some extent on
DWBA calculations which, as we have shown, do not account properly
for the measured data. Consequently, the experimental data for 
$(d,p_0)$ and $(d,n_0)$
reactions were not included in our searching procedure, and the spectroscopic
factors involved in these transitions were set to unity.

\begin{table}
\caption{\label{Tab:CRC}Deuteron optical potential parameters and spectroscopic
amplitudes resulting from CRC calculations performed for 
the $^{16}\text{O}(d,n_{1})^{17}\text{F}$and 
$^{16}\text{O}(d,p_{1})^{17}\text{O}$  angular distribution at $E_{d}=2.85$ MeV. The 
parameters not listed in the table are those of potential (a) from
Table \ref{Tab:OP1}.}

\begin{ruledtabular}
{\centering \begin{tabular}{cccccc}
&
 $V_{0}$ &
 $W_{d}$ &
 $a_{i}$ &
\multicolumn{2}{c}{Spectroscopic Amplitude}\\
&
(MeV)&
(MeV)&
(fm)&
$\langle ^{17}\text{F}^{*}|^{16}\text{O}\rangle $ &
$\langle ^{17}\text{O}^{*}|^{16}\text{O}\rangle $ \\
\hline 
Set I  & 102  & 20.2  & 0.232 & 1.00$^a$ & 1.00$^a$ \\
Set II & 104  & 24.9  & 0.233 & 0.89$\pm $0.02 & 0.95$\pm $0.01 \\
\end{tabular}\par}
\end{ruledtabular}
$^a$ These values where kept fixed in this search.
\end{table}

The best fit parameters corresponding
to different searches are presented in Table \ref{Tab:CRC}. 
For the set I, only the depths and the diffuseness of
the imaginary part
of the deuteron 
central potential were considered as free parameters.
These parameters were adjusted as to minimize the $\chi ^{2}$
for the $(d,d)$ angular distribution. The radii and the 
real part diffuseness are the same as in Table \ref{Tab:OP1}. All 
spectroscopic amplitudes were set to one. 

As one can see, the imaginary part of the $d+^{16}$O 
optical potential, which comes out from the CRC analysis, is 
much deeper and less diffuse than the deuteron optical potential (a)
listed in Table I.

\begin{figure}
{\centering \resizebox*{0.95\columnwidth}{!}{\includegraphics{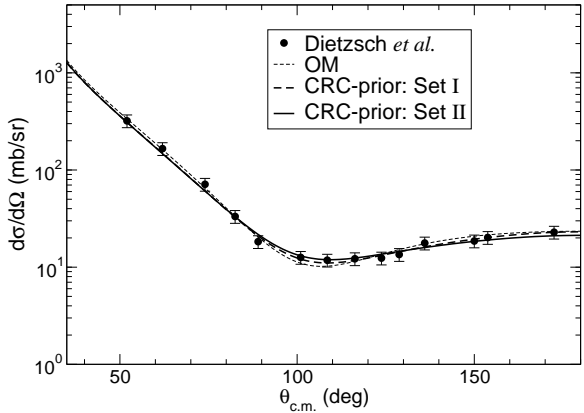}} \par}

\caption{\label{Fig:dd_e285}Experimental and calculated 
elastic angular distributions for $d+^{16}$O at $E_d$=2.85 MeV. }
\end{figure}

\begin{figure}
{\centering \resizebox*{0.95\columnwidth}{!}
{\includegraphics{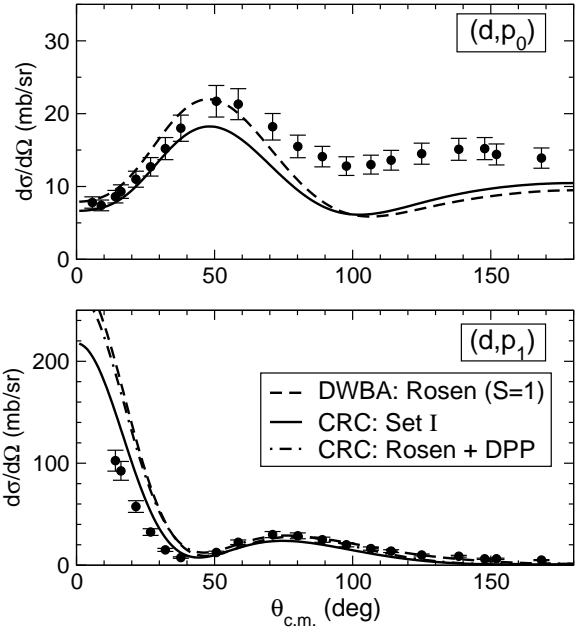}} \par}

\caption{\label{Fig:CRC1_dp}CRC calculations for the 
$^{16}\text{O}(d,p)^{17}\text{O}$
reaction at $E_d$=2.85 MeV, using the set I of 
parameters (see Table \ref{Tab:CRC}).
The contributions of 2.95 mb/sr and 1.14 mb/sr, coming from 
CN formation, have been added
to the $(d,p_0)$ and $(d,p_1)$ distributions, respectively. }
\end{figure}

\begin{figure}
{\centering \resizebox*{0.95\columnwidth}{!}
{\includegraphics{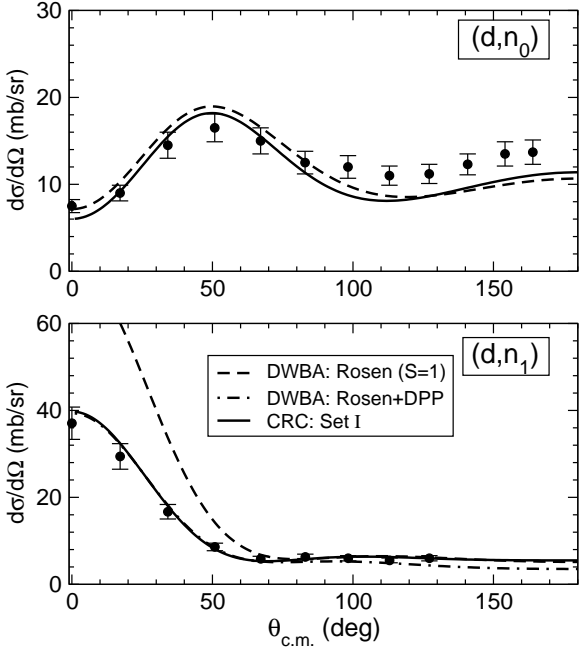}} \par}

\caption{\label{Fig:CRC1_dn}CRC calculations for the
$^{16}\text{O}(d,n)^{17}\text{F}$ reaction
at $E_d$=2.85 MeV, 
using the set I of parameters (see Table \ref{Tab:CRC}).
The $(d,n_{0}$) includes a contribution of 2.0~mb/sr, 
coming from CN formation.}
\end{figure}

The CRC calculations for the 
$(d,p)$ and $(d,n)$ angular distributions, using the set I of parameters,
are presented
in Figs.~\ref{Fig:dd_e285}, \ref{Fig:CRC1_dp}, and \ref{Fig:CRC1_dn},
respectively. For comparison purposes, the DWBA prediction, assuming
unit spectroscopic factor, is also included in the figure. The
CRC $(d,d)$ distribution  (thick dashed line in Fig.~\ref{Fig:dd_e285})
is in perfect agreement with the data. Also, these calculations
preserve the agreement with the $(d,n_0)$ and $(d,p_0)$ distributions,
as compared with the DWBA calculations.
Furthermore, the CRC calculations produce a  reduction 
in the cross section at forward angles for  
the $(d,p_1)$ and $(d,n_1)$ reactions, improving 
significantly the agreement with the experimental with
spectroscopic factors close to one. 

\begin{figure}
{\centering \resizebox*{0.95\columnwidth}{!}
{\includegraphics{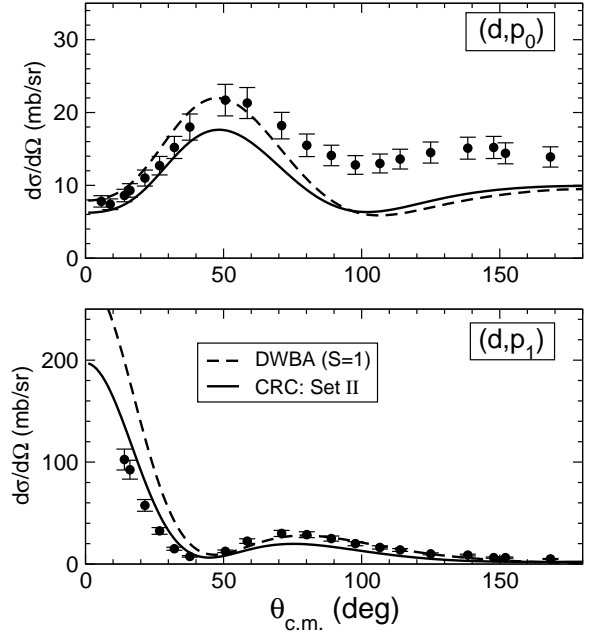}} \par}
\caption{\label{Fig:CRC2_dp}CRC calculations for the $(d,p_0)$ 
and $(d,p_1)$ channels in the ${d+}^{16}\text{O}$ reaction at $E_d$=2.85 MeV,
using the set II of parameters (see Table \ref{Tab:CRC}), for the
incoming distorted potential. The $(d,p_{0})$ and $(d,p_{1})$ include 
the contributions of 2.95 
and 1.14 mb/sr, respectively, coming from CN formation. }
\end{figure}

\begin{figure}
{\centering \resizebox*{0.95\columnwidth}{!}
{\includegraphics{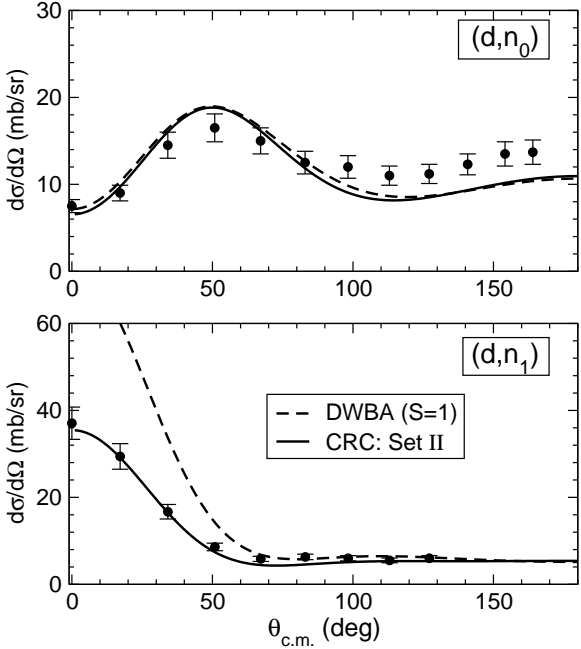}} \par}
\caption{\label{Fig:CRC2_dn}CRC calculations for the $(d,n)$
channels 
in the $d+^{16}\text{O}$ reaction at $E_d$=2.85 MeV,
using the set II of parameters (see Table \ref{Tab:CRC}). }
\end{figure}

In a second search, set II in the Table II, $V_{0}$, $W_{d}$, $a_{i}$
and the spectroscopic amplitudes for 
the overlaps $\langle ^{17}\text{O}^{*}|^{16}\text{O}\rangle$
and $\langle ^{17}\text{F}^{*}|^{16}\text{O}\rangle $ were
set as free parameters.
As a result of the $\chi ^{2}$ minimization, the real and imaginary
depths were slightly modified with respect to the values of the previous
search, while the imaginary diffuseness results also on a small
value. Interestingly, the extracted spectroscopic factors are very close
to 1. The results of this search, which are presented in 
Figs.~\ref{Fig:dd_e285}, \ref{Fig:CRC2_dp}
and \ref{Fig:CRC2_dn}, are very similar
to those obtained in the previous fit, the main difference being a
slight improvement in the fit for the $(d,n_1)$ distribution 
at forward angles. The  calculated angular distributions for the 
$(d,n_0)$ and $(d,p_0)$ reactions agree very well with the data, 
although the backward angular region is still underestimated. 
The calculation, also, overestimates the $(d,p_1)$ distribution.

As it has been said before, the extracted deuteron potential is
less diffuse than the OP(a).
This suggests that the  imaginary part of this bare potential  
comes from a short-range process, such as compound nucleus 
formation. This interpretation
is consistent with the fact that, in our CRC calculations, all the
relevant direct couplings are explicitly included. Note that at these
scattering energies, target excitation is forbidden by energy conservation,
and projectile breakup is expected to be very small due to the restricted
phase-space available. Therefore, the only channels that could contribute
to the absorption of the $d+^{16}$O potential, besides those already
included, are those
leading to compound nucleus formation and, possibly, a small direct
contribution coming from the $(d,\alpha )$ process.
In this respect we note that the experimental excitation functions 
for the transfer reactions, 
in this energy region, exhibit structures which, in principle, 
could be due  to a reminiscent effect of the resonances in the
compound nucleus (see, for instance, Ref.~\cite{Die68}). 
Near these resonances, the meaning and usefulness of the optical 
model is questionable and all the conclusions reached above
can be attributed to an inadequacy of the DWBA calculation.

To rule this possibility out, we have extended our analysis to 
other energies, ranging from $E_d$= 2.29 to $E_d$= 3.186 MeV for $(d,n)$ 
reaction and from $E_d$= 2.279 to $E_d$= 3.155 MeV for the $(d,p)$ reaction.
The average spectroscopic factors obtained from the CRC analysis are 
summarized 
in Table \ref{Tab:cfp} for two different sets of the incoming channel
optical potential. The standard deviation specified as the error of the 
spectroscopic factors were obtained from the average of the four 
energies analyzed.
Except for the $(d,p_1)$ channel, all spectroscopic factors are close to one.
The small value found for the $\langle^{17}\text{O}^*|^{16}\text{O}\rangle$
spectroscopic factor
should be considered an open problem in our analysis.

\begin{table}
\caption{\label{Tab:cfp} Extracted values for the spectroscopic factors 
          derived from CRC calculations. }
\begin{ruledtabular}
{\centering \begin{tabular}{ccccc}
&\multicolumn{4}{c}{Average spectroscopic factor} \\
&$(d,p_0)$ &  $(d,p_1)$ &  $(d,n_0)$  & $(d,n_1)$  \\
\hline

Set I  &   1.14(7)  &  0.70(16) &  0.97(11)  &   1.00(12)    \\ 

Set II &   1.19(5) &  0.69(17) &  0.93(11)  &   0.96(12) \\

\end{tabular} \par}
\end{ruledtabular}
\end{table}

\section{DISCUSSION}\label{sec:discussion}

The large discrepancy between the CRC and DWBA calculations presented
above suggests that the Born approximation may not be valid to 
describe the angular distribution of the studied
transfer reactions. This conclusion might depend, nevertheless,
on how the potentials in the DWBA amplitude are defined. In
the standard DWBA, the entrance and exit distorting potentials
are defined as to reproduce the elastic scattering in their respective
channels. In our case, the entrance optical potential could be 
determined accurately, since experimental data for the elastic
channel was measured in the experiment of Dietzsch
{\it et al} \cite{Die68}.
However, for the ${n+}^{17}$F exit channel, 
no elastic data exist due to the exotic nature of the $^{17}$F
nucleus. For the ${p+}^{17}$O system, the only low energy
data available in the literature, up to our knowledge, consist on 
excitation functions
for elastic scattering in the energy range $E_{p}=$0.5-1.33~MeV
\cite{Sen73} and $E_{p}=$1.4-3~MeV \cite{Sen77} for a few
scattering angles. Thus, to generate the distorted waves for the 
exit channel in our analysis of the  $(d,p_{0})$ reaction, we rely on the 
Rosen parameterization, which  reproduces
reasonably well the data of Ref. \cite{Sen77}. 
However, there is no guarantee that this OP  describes properly 
the (hypothetical) elastic scattering for ${p+}^{17}\text{O}^{*}$ system, 
where the 
target  is in the first excited state. 
Actually, the CRC calculations, presented 
throughout this work, clearly indicate that the $d \rightarrow p_{0}$ and 
$d \rightarrow p_{1}$
couplings have very different strengths, the latter being much stronger.
Thus, as shown below, different optical potentials
were required for ${p+}^{17}\text{O}_{gs}$ and ${p+}^{17}\text{O}^{*}$.
A similar argument and conclusion holds for the OP for ${n+}^{17}\text{F}_{gs}$
and ${n+}^{17}\text{F}^{*}$ system.

\begin{figure}
\resizebox*{0.97\columnwidth}{!}{\includegraphics{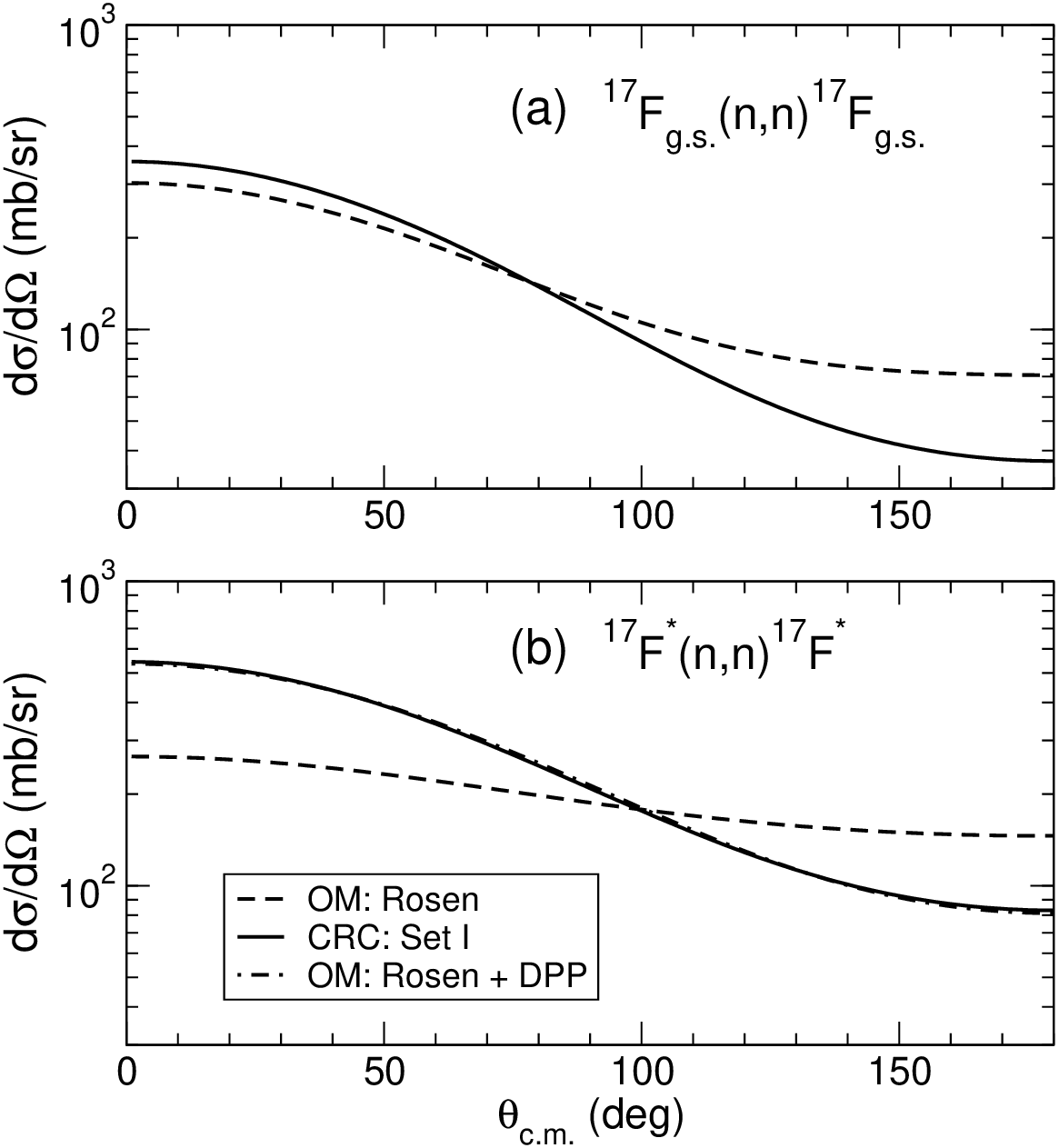}}
\caption{\label{Fig:f17nn}Elastic scattering for ${n+}^{17}$F,
at $E_{n}$=0.9~MeV and $E_{n}$=0.4~MeV,
with $^{17}$F initially in the ground (upper panel)
or excited (lower panel) state, respectively.}
\end{figure}

To get further insight into this problem,
OM elastic scattering calculations obtained with the Rosen parameterization 
were compared with the result of the CRC calculation for ${n+}^{17}\text{F}_{gs}$
and ${n+}^{17}\text{F}^{*}$ at the neutron energy appropriate
for the $(d,n)$ reaction at $E_{d}$=2.85~MeV. These calculations 
are shown in Fig.~\ref{Fig:f17nn}.
The dashed lines are the OM calculations with 
the Rosen parameterization and the solid
lines are the CRC calculation using the parameters from set I (see Table
\ref{Tab:CRC}). As it can be seen, both ${n+}^{17}\text{F}_{gs}$
and ${n+}^{17}\text{F}^{*}$ angular distributions are clearly modified
when coupling to the transfer channels are included. Interestingly, the
$^{17}\text{F}(n_{0},n_{0})^{17}\text{F}$ elastic 
scattering remains basically unchanged
at small angles. This result might explain why the 
DWBA calculation reproduces the 
angular distribution for the $^{16}\text{O}(d,n_{0})^{17}\text{F}$ 
transfer and not the
$^{16}\text{O}(d,n_{1})^{17}\text{F}$ channel. On the other hand, the 
$^{17}\text{F}(n_{1},n_{1})^{17}\text{F}$ scattering is strongly 
 enhanced at forward angles due to the coupling
to the transfer channels. It becomes apparent that an OP that fits
the $(n_{0},n_{0})$ elastic scattering will not reproduce the $(n_{1},n_{1})$
scattering. Consequently, this OP will not be suitable as distorted
potential for the DWBA amplitude of the $(d,n_{1}$) process.

One could go further and ask whether an optical potential
that fits the $(n_1,n_1)$ elastic scattering angular distribution, 
given by the CRC calculation, could be used as distorting potential 
in the DWBA amplitude to improve the agreement of the DWBA
calculations for the $(d,n_{1})$ angular distribution with the data. 
To answer this question, 
a phenomenological OP has been added to the one
obtained with the Rosen parameterization in a such way 
that the combined potential reproduces the $(n_1,n_1)$
angular distribution given by the CRC calculation.

Obviously, the choice of this extra potential is not unique and, 
for simplicity, just the imaginary part, with a surface Woods-Saxon 
shape, has been considered. The extracted
parameters are listed in Table~\ref{Tab:OP1}, set (d). 
The corresponding calculated angular
distribution is shown  in Fig.~\ref{Fig:f17nn}(b) 
indicated by the dotted-dashed line. 
As it can be seen, the CRC effects are perfectly accounted 
for  by using this phenomenological OP. Furthermore, by using this 
potential as a distorting potential for the exit channel in the $(d,n_{1})$ reaction, 
the DWBA calculated angular distribution is in excellent agreement with the CRC 
calculation and, hence, with the experimental data. 
The result of this calculation is shown in the bottom panel
of Fig.~\ref{Fig:CRC1_dn}, indicated by the dotted-dashed line
(Rosen+DPP). This result suggests that the additional potential can be 
regarded as a \emph{dynamic polarization potential} (DPP) that accounts 
for the coupling effect of the $(d,n_1)$ channel in the 
$^{17}\text{F}(n_{1},n_{1})^{17}\text{F}$ elastic
scattering. 

\begin{figure}
\resizebox*{0.97\columnwidth}{!}{\includegraphics{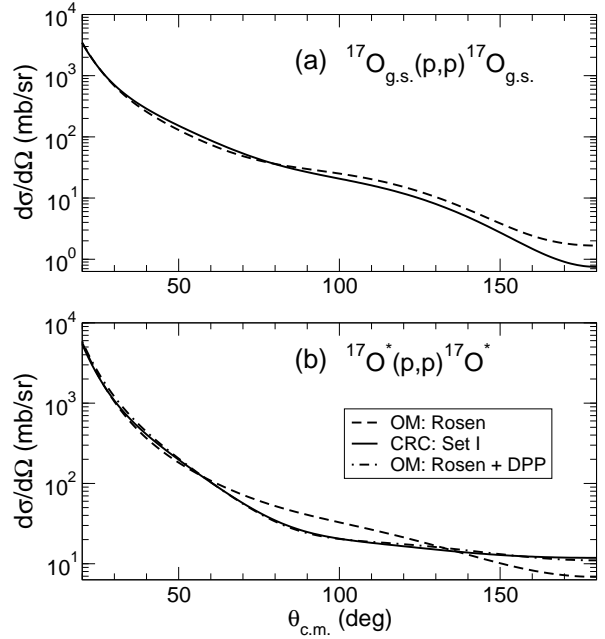}}
\caption{\label{Fig:o17pp}Elastic scattering for ${p+}^{17}\text{O}$,
at $E_{p}$=4.7~MeV and $E_{n}$=3.8~MeV,
with $^{17}$O initially in the ground (upper panel)
or excited (lower panel) state, respectively.}
\end{figure}

A similar analysis was carried out for the proton channel. 
The calculated angular distributions for the ${p+}^{17}\text{O}$
elastic scattering,  at the outgoing proton energy for the 
reaction $^{16}\text{O}(d,p)^{17}\text{O}$ at $E_d$=2.85 MeV, are shown 
in Fig.~\ref{Fig:o17pp}. Again, the difference between the pure
optical model calculation (dashed lines) and the CRC calculation (solid
lines) is more pronounced in the $(p_1,p_1)$ case than in the 
$(p_0,p_0)$ case. This is a clear indication that the $d \rightarrow p_1$ coupling
is stronger than the $d \rightarrow p_0$ one. 
In analogy with the neutron case, a DPP potential has also been added to 
the ${p+}^{17}$O potential to reproduce the CRC elastic scattering
distribution for $(p_1,p_1)$. In this
case, a complex potential, comprising a  real volume term and a imaginary surface
part, has been used.
The parameters for this potential are listed in 
Table \ref{Tab:OP1}, set (e). As it can be seen, in this case
the polarization potential is repulsive and has
a diffuse and shallow absorptive component.

Unlike  the neutron case, by using
this extra polarization potential together with the potential (c) 
as distorting potential in the DWBA calculation, did not 
reproduce the CRC result for the $^{16}\text{O}(d,p_1)^{17}\text{O}$ 
reaction. The result of the DWBA calculation with this extra DPP, 
shown in  Fig.~\ref{Fig:CRC1_dp} by the dotted-dashed line, is very similar to the
DWBA calculation with the bare potential alone. Of course, since 
the extra DPP potential is not unique, there is always the 
possibility of that another more appropriate DPP would improve the 
agreement with the experimental angular distribution for the  
$^{16}\text{O}(d,p_{1})^{17}\text{O}$ reaction. 
Unfortunately, we have not been able to find such a potential.  
Notwithstanding these considerations, we 
would like to stress that it is not obvious that the (non-local) transfer coupling
can be described in general by a simple local  potential. 
To support this conclusion, we have   
calculated the trivially local equivalent polarization potential  \cite{Coul77} 
for the $p+^{17}\text{O}^*$ and $n+^{17}\text{F}^*$ elastic scattering, using the solution 
provided by our CRC calculations. The polarization potential so obtained was found
to be very oscillatory, and strongly $L-$dependent, supporting the idea that 
transfer couplings are not easily representable by simple Woods-Saxon forms.
A similar analysis was performed by Coulter and Satchler \cite{Coul77}, 
reaching similar conclusions.

It has been argued  by several authors  \cite{Sat66b,Asc78} that in 
some cases the appropriate incoming (exit)
distorting potential to be used in the DWBA amplitude does not necessarily 
fits the experimental elastic scattering  in the  entrance (exit) channel.
Instead, these authors suggest the use of an alternative
prescription in which the distorted potential is replaced by the 
bare potential, as obtained from a CC or CRC calculation.
Ichimura and Kawai \cite{Ich83}, for instance, have investigated the validity of the
conventional and alternative expressions of the DWBA amplitude for the 
$^{16}\text{O}(d,p_{1})^{17}\text{O}$ transfer reaction.
However, the  found that both DWBA prescriptions fail to reproduce the CRC result.
Our calculations seem to support this conclusion. 

\section{Summary and conclusions}\label{sec:summary}

In this work we have studied the  $^{16}\text{O}(d,p)^{17}\text{O}$ and
$^{16}\text{O}(d,n)^{17}\text{F}$ transfer reactions at sub-Coulomb 
energies ($E_d \approx 2-3$ MeV).
We have shown that  standard DWBA calculations, that satisfactory reproduce 
the $(d,n_0)$ and $(d,p_0)$ forward angular distributions, do not
quite reproduce the $(d,n_1)$ and $(d,p_1)$ data, unless anomalously
small spectroscopic
factors are used for the $\langle ^{17}\text{F}^*|^{16}\text{O}\rangle$
and $\langle ^{17}\text{O}^*|^{16}\text{O}\rangle$ overlaps. This discrepancy
remains even when couplings between excited states of the final nucleus 
are included through the CCBA formalism. A full coupled reaction channels (CRC)
calculation, which treats the transfer couplings 
beyond the Born approximation, greatly improves the agreement 
with the data. In  particular, the polarization between the entrance $(d,d)$ 
channel and the $(d,p_1)$ and $(d,n_1)$ reaction channels reduces the cross 
sections at forward angles, resulting in a very
good agreement with the data, while maintaining spectroscopic 
factors close to one, particularly in the case of the $(d,n_1)$ reaction.
In that case, we have found that these higher order effects
can be accounted for
within the DWBA formalism by adding an effective optical 
local potential to the
exit channel distorting potential for $n+^{17}$F$^{*}$. 
This polarization  potential is chosen in a such way that the total 
(bare + DPP) potential reproduces the elastic data
on the exit channel. Therefore, in this particular
case, one can still use the DWBA formalism, provided that different distorting 
potentials are used for the $(d,n_0)$ and $(d,n_1)$ channels. Unfortunately, 
such a DPP could not be found for the $(d,p_1)$ channel. Other authors \cite{Ich83} were 
also not able to find such DPP for the same reaction at  higher 
energies. 
With the  present analysis we aim to call the attention to 
some of the limitations in the DWBA calculation as a tool to extract 
spectroscopic information from reactions of astrophysical interest.

\begin{acknowledgments}
The authors would like to thank FAPESP for the financial support 
(Projeto tem\'atico 01/06676-9). 
A.M.M. acknowledges a postdoctoral grant by the Funda\c c\~ao para a Ciencia
e a Tecnologia (Portugal). We acknowledge useful discussions with 
J. G\'omez Camacho. 
\end{acknowledgments}
\bibliographystyle{apsrev}
\bibliography{./o16d}
\end{document}